# A Federated Learning Framework for Privacy-Preserving Oral Cancer Screening on Smartphones

Lena D. Swamikannan, Akshay Bhagwan Sonawane, Jay S. Patel, C.S. Mani, Lakshmi Narayana, and Lakshman Tamil, *Senior Member, IEEE*

***Abstract* — Data are the cornerstone of robust AI models. However, in the medical domain, access to reliable data is constrained by regulatory requirements and patient privacy, and clinical oral images are particularly difficult to obtain. Federated learning (FL) mitigates these constraints by enabling collaborative model development across decentralized datasets without centralizing or sharing patient data.**

**This work presents a practical FL framework that supports geographically distributed collaboration among AI healthcare researchers and facilitates the development of robust models for oral cancer screening. Client devices were interconnected via Tailscale to provide secure networking and real-time communication. We implemented the FL workflow using the Flower framework for server-side aggregation, while client deployment and orchestration were configured manually; no enterprise FL platforms were used. To support a smartphone-based screening application, we evaluated lightweight, mobile-friendly architectures including MobileNetV2, MobileNetV3Large, and MobileNetV4-Conv-Small (MNv4-Conv-S). Across the global lightweight models aggregated using FedAvg, the MNv4-Conv-S based global model (GM-V4) achieved the best performance, reaching an AUC of 0.929 and an accuracy of 87%.**

## I. Introduction

ORal cancer (oral cavity cancer) is a major subtype of head and neck malignancies. In U.S. cancer registry reporting, oral cavity and oropharyngeal cancers were grouped together starting in 1973; however, they are now widely recognized as biologically and clinically distinct diseases. This distinction became increasingly evident in the mid-2000s and was formally reinforced in the American Joint Committee on Cancer (AJCC) 8th edition staging system (2017).

Oral cavity cancer primarily involves anterior oral structures (e.g., lips, buccal mucosa, floor of mouth, anterior two-thirds of the tongue), whereas oropharyngeal cancer arises in the posterior pharyngeal region (e.g., base of tongue, tonsillar region, and soft palate). One of the most common oral cancer sites globally is the lateral border of the tongue. Clinically, oral cancers are staged using the International Union Against Cancer (UICC) TNM (Tumor, Node, Metastasis) system, which guides treatment and prognosis. Early detection is strongly associated with reduced mortality, less aggressive treatment, less disability, better quality of life, lower financial and psychosocial burden.

In the United States, the American Cancer Society [1] estimates that in 2026 there will be approximately 60,480 new cases of oral cancer and 13,150 deaths. Two complementary lines of evidence motivate earlier detection. First, a randomized controlled trial (RCT) of oral cancer screening in a high-risk population demonstrated improved outcomes with systematic screening [2]. Second, population-level SEER (Surveillance, Epidemiology, and End Results) summary staging data [3] consistently show substantially higher survival when disease is detected at localized stages.

The Kerala oral cancer screening RCT was conducted in a high-risk community population over nine years (1996–2004). Thirteen clusters were formed: seven intervention clusters (oral screening) and six control clusters (standard care). Two non-medical health workers were assigned per cluster; intervention workers were trained to perform oral visual inspection using standardized manuals with clinical descriptions and color photographs of oral lesions. The intervention arm achieved statistically significant improvements in outcomes [2].

In SEER reporting, oral cavity and pharynx cancers are commonly summarized by SEER Summary Stage (localized, regional, distant). Across diagnosis years (Fig. 1), localized disease shows the highest 5-year relative survival (85–90%), regional disease is intermediate (65–70%), and distant disease remains the lowest (35–45%), underscoring the survival advantage of early-stage detection [3].

Despite rapid progress in medical AI, building robust oral cancer screening models remains constrained by limited access to diverse, well-annotated clinical oral images. Privacy regulations, the need for specialist labeling, and site-specific variation in demographics and imaging conditions make centralized dataset creation difficult, and models trained at a single site often generalize poorly. Federated learning (FL) offers a

Lena D. Swamikannan (lxd200013@utdallas.edu), Akshay B. Sonawane (axs180315@utdallas.edu) and Lakshman S. Tamil (laxman@utdallas.edu) are with Erik Jonsson School of Engineering and Computer Science, The University of Texas at Dallas, Richardson, USA.
Jay S. Patel (patel.jay@temple.edu) Department of Oral Health Sciences,Temple University Kornberg School of Dentistry,Philadelphia,USA
C.S.Mani(dr.cs.mani@gmail.com) and Lakshmi Narayana (drnarayana777@gmail.com) are with Apollo Hospital, Chennai, INDIA.

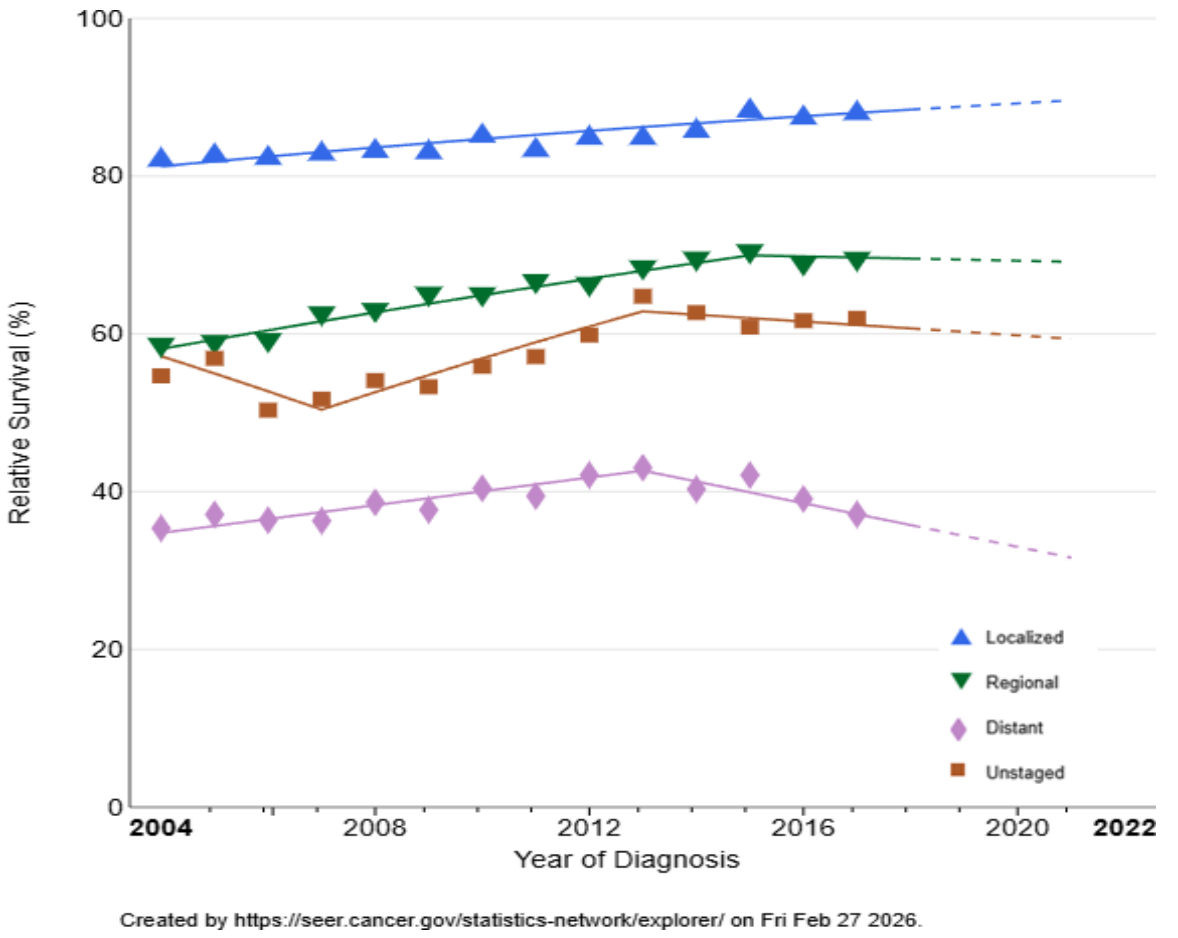

Fig. 1. Relationship between SEER summary stage at diagnosis (localized, regional, distant) and 5-year relative survival for oral cavity and pharynx cancer, highlighting the survival advantage of localized-stage diagnosis [3]. A solid trend line represents the predicted (modeled) survival trend. A dashed trend line represents the projected survival trend.

practical alternative by enabling collaborative training across institutions without centralizing raw patient data. A detailed analysis of security attacks, privacy leakage, and adversarial threats is beyond the scope of this work and left for future study.

FL is therefore well-suited for privacy-preserving AI in oral healthcare, supporting collaborative learning across geographically distributed data sources while helping maintain compliance with healthcare privacy requirements (e.g., HIPAA [4]). Moreover, deploying trained global models to edge devices such as smartphones can enable real-time triaging and screening, potentially improving access to early detection and reducing delays in referral pathways.

**Gap and novelty:** Prior work on smartphone-based oral cancer screening has largely assumed centralized training and single-site data access, while FL studies in healthcare have focused predominantly on radiology and structured EHR modalities rather than photographic oral cavity images. As a result, there is limited evidence on how to build and operate an end-to-end, cross-site FL pipeline for clinical oral photographs and then deploy the resulting global model for on-device inference. This paper addresses that gap by presenting a practical, real-time FL framework tailored to oral cancer screening with smartphone-compatible models and a working mobile deployment.

The main contributions of this paper are:

- We present an end-to-end, real-time cross-site FL pipeline for oral cancer screening trained on decentralized clinical oral photograph datasets.
- We implement a low-overhead, secure FL testbed using Tailscale and the Flower framework to enable decentralized training across geographically separated sites (Philadelphia and Dallas).
- We curate and introduce customized oral image datasets (SmartOralpix and Philly-Oral) to support federated evaluation.
- We develop and compare federated models using lightweight pretrained architectures suitable for mobile deployment (MobileNetV2, MobileNetV3Large, and MobileNetV4-Conv-Small).
- We develop an iOS smartphone application for oral cancer screening and integrate the best-performing global model to enable on-device inference.

The remainder of this paper is organized as follows. Section II reviews related work on AI-based analysis of clinical oral images and federated learning. Section III describes the datasets used in this study. Section IV presents the proposed federated learning framework and training methodology. Section V reports experimental results and provides discussion. Section VI outlines the design and implementation of the smartphone application. Finally, Section VII concludes the paper and summarizes limitations and directions for future work.

## II. Related Work

Most deep-learning work on oral cancer detection has focused on *oral histopathology* images trained in centralized settings [17]–[21]. A key reason is the availability of public histopathology datasets. Because these images are obtained after biopsy, such systems primarily assist pathologists and support diagnosis rather than frontline screening.

In contrast, this study targets *clinical photographic oral cavity images* and emphasizes privacy-preserving training suitable for smartphone deployment. Table I summarizes prior work using clinical oral images, the backbone architectures explored, and the training setting with respect to data centralization.

### A. AI Models for Clinical Oral Images

Deep learning for medical image analysis is dominated by a small number of architecture families, including convolutional neural networks (CNNs), Transformers, recurrent neural networks (RNNs), and hybrid designs (e.g., CNN+Transformer). For deployment, models are often categorized as *lightweight* or *heavyweight* based on parameter count, memory footprint, latency, and power consumption. Table II summarizes typical trade-offs.

Because our goal is on-device inference for smartphone screening, we focus on lightweight backbones. Specifically, we evaluate MobileNetV2 [14], MobileNetV3Large [15], and MNv4-Conv-S [16], all pretrained on ImageNet.

### B. Federated Learning in Healthcare and Oral Cancer

Federated learning (FL) enables multiple clients (e.g., hospitals, laboratories, or devices) to collaboratively train a global model while keeping raw data local. This paradigm supports data ownership and helps address privacy and compliance constraints in healthcare.

A representative multi-site healthcare FL study is reported by Peng *et al.* [22], who trained a COVID-19 chest radiograph model across sites in the United States and Europe using NVIDIA Clara Train SDK 4.0. They evaluated several aggregation variants (FedAvg, FedBN, FedProx, and FedAMP) with a DenseNet-121 backbone. In oral cancer, evidence remains comparatively limited. Firdaus *et al.* [13] proposed an FL framework for oral cancer detection using clinical images with a DenseNet-201 backbone and compared FedAvg, FedProx, and centralized training; FedAvg yielded the best accuracy (97.48%). Rajit *et al.* [23] studied FL on an oral histopathology dataset and analyzed IID versus non-IID partitions using a CNN-based model, reporting 90.18% accuracy for the FedAvg global model versus 99.65% under centralized training. Together, these studies motivate FL for privacy-preserving collaboration and highlight the need for practical approaches tailored to clinical oral photographs and mobile deployment.

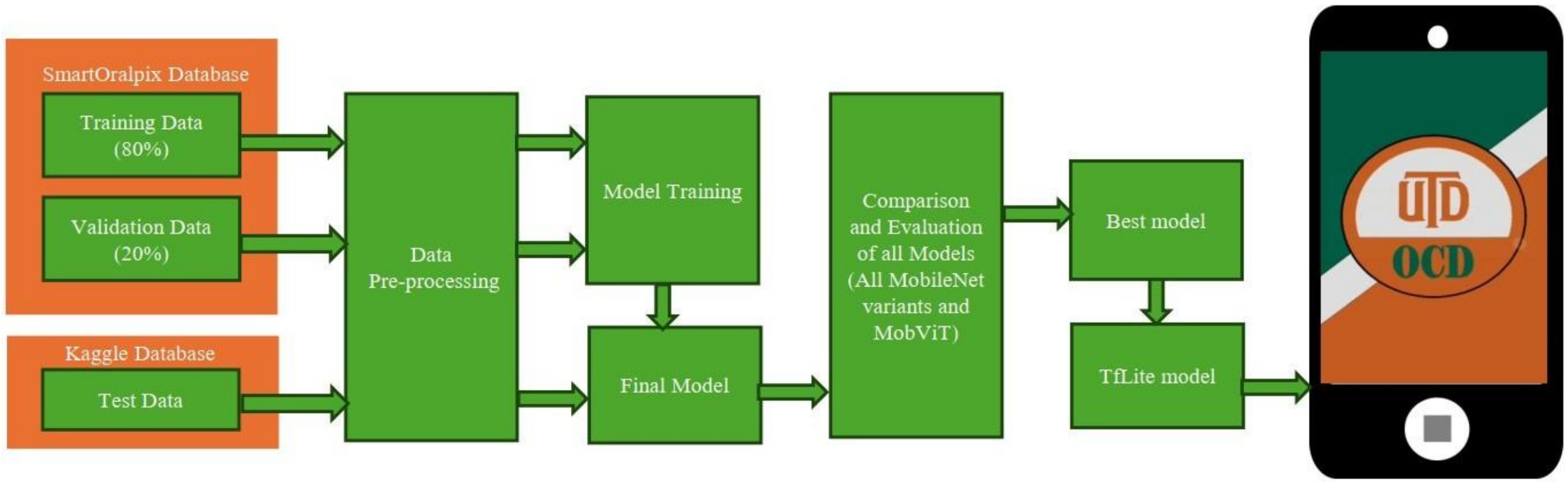

Fig. 2. Workflow for developing a telemedical smartphone application for oral cancer screening as presented by Lena *et al.* [5]. Our approach augments this workflow with a federated learning framework to enable cross-site collaboration and improve generalization under demographic and clinical variability.

TABLE I
SUMMARY OF RESEARCH STUDIES IN AUTOMATED ORAL CANCER DETECTION

| Reference | Data Modality | Architecture Explored | Source | Privacy |
|---|---|---|---|---|
| [6] | Clinical Oral Images | Heavyweight CNN (Inception-ResNet-V2) | Private | ✗ |
| [7] | Clinical Oral Images | Heavyweight CNN (DenseNet121, Faster R-CNN) | Private | ✗ |
| [8] | Clinical Oral Images | Lightweight CNN (MobileNetV2) and Heavyweight CNN (DenseNet, ResNet50, VGG19, VGG16) | Private | ✗ |
| [9] | Clinical Oral Images | Heavyweight CNN (ResNet101, Faster R-CNN) | Private | ✗ |
| [5] | Clinical Oral Images | Lightweight CNN (MobileNet (V1, V2, V3Small, V3Large) and Lightweight Hybrid (MobileViT-S) | Private | ✗ |
| [10] | Clinical Oral Images | Lightweight CNN (EfficientNet-B0) | Private | ✗ |
| [11] | Clinical Oral Images | Lightweight CNN (MobileNet) and Heavyweight CNN (VGG19, VGG16, Inceptionv3, ResNet50) | Private | ✗ |
| [12] | Clinical Oral Images | Heavyweight Transformer (Swin-Transformer) | Private | ✗ |
| [13] | Clinical Oral Images | Heavyweight CNN (DenseNet-201) | Public | ✓ |
| **This study** | **Clinical Oral Images** | **Lightweight CNN (MobileNetV2, MobileNetV3Large, MNv4-Conv-S)** | **Private** | ✓ |

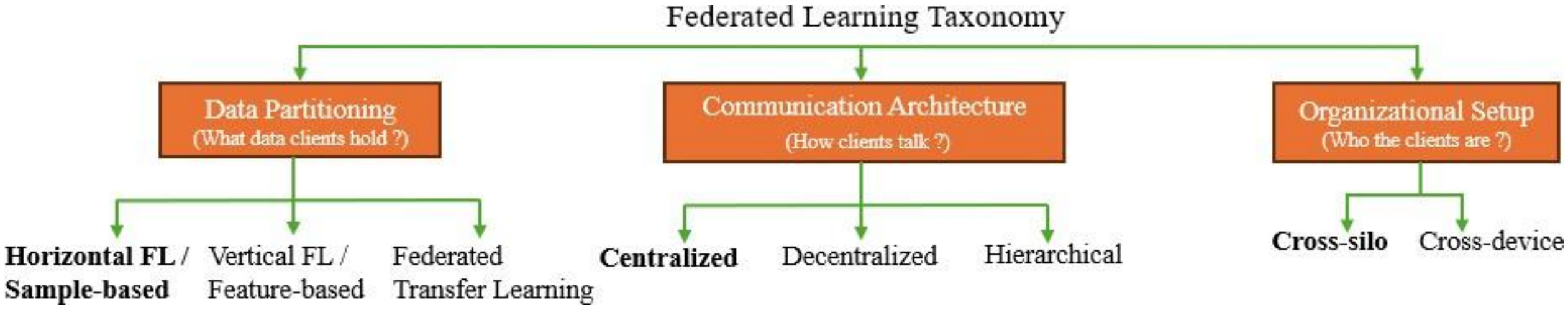

Fig. 3. Taxonomy of federated learning (FL) based on Data distribution, Communication Architecture, and Organizational setup.

TABLE II
LIGHTWEIGHT VS HEAVYWEIGHT MODELS

| Factors | Lightweight Models | Heavyweight Models |
|---|---|---|
| Model size | Small | Large |
| Inference | Fast | Slow |
| Accuracy | Slightly Lower | High |
| Usage | Realtime | Offline |
| Deployment | Mobile, Edge Device | Cloud deployment |

### C. Federated Learning Taxonomy

FL can be categorized along three axes: data partitioning, communication architecture, and organizational setup [24], [25]. Fig. 3 illustrates the structure of the FL taxonomy.

TABLE III
SUMMARY OF MOBILE-FRIENDLY MODELS UTILIZED IN THIS STUDY

| Models | Architecture | Input image size (W × H × D) | No.of parameters (millions) |
| --- | --- | --- | --- |
| MobileNetV2 [14] | CNN | 224 × 224 × 3 | 3.4 |
| MobileNetV3Large [15] | CNN | 224 × 224 × 3 | 5.4 |
| MNv4-Conv-S [16] | CNN | 224 × 224 × 3 | 3.8 |

*1) Data Partitioning:*

- **Horizontal FL (HFL):** clients share the same feature space but hold different samples. Local inference is feasible.
- **Vertical FL (VFL):** clients share samples but have different feature spaces. Local inference typically requires collaboration.
- **Federated Transfer Learning (FTL):** both feature and sample spaces differ, relying on transfer learning techniques.

*2) Communication Architecture:*

- **Centralized:** hub-and-spoke topology in which clients communicate only with a central aggregator.
- **Decentralized:** peer-to-peer topology in which clients exchange updates directly without a central authority.
- **Hierarchical:** multi-tier topology with edge/regional aggregators that combine client updates and coordinate with a global server.

*3) Organizational Setup:*

- **Cross-silo:** a small number of reliable clients (e.g., hospitals, universities) with stable connectivity.
- **Cross-device:** a large number of edge devices (e.g., smartphones, wearables) with intermittent connectivity and limited resources.

Based on this taxonomy, our setup corresponds to *horizontal FL* with a *centralized* communication architecture in a *cross-silo* organizational setting.

### D. Federated Learning Challenges

Client heterogeneity is a central challenge in FL, and prior work has proposed algorithmic and systems-level approaches to address it [26], [27]. Broadly, heterogeneity can be grouped into statistical heterogeneity and model heterogeneity.

*1) Statistical Heterogeneity:* Statistical heterogeneity (non-IID data) occurs when clients hold data drawn from different distributions. It can arise as label skew, quantity skew, quality skew, or feature skew. In clinical practice, label skew is common when some sites contain few malignant cases relative to normal cases. Feature skew can arise due to demographic differences, acquisition protocols, lighting, or camera hardware, which shift feature distributions across sites.

*2) Model Heterogeneity:* Model heterogeneity occurs when clients use different architectures or capacities due to differences in compute resources. This setting is often addressed via knowledge-distillation-based FL, where constrained clients train smaller student models while more capable clients or the server maintain larger teacher models.

## III. MATERIALS AND METHODS

### A. Motivation

In our previous work [5], we developed a telemedical smartphone application for oral cancer screening by fine-tuning five lightweight models on the SmartOralpix dataset under centralized training. Using the same test protocol adopted in this paper, the best-performing model (MobileNetV3Large) achieved 84% accuracy and was deployed in an Android application. However, SmartOralpix is relatively small, and models trained on a single-site dataset may not generalize well to real-world screening conditions.

Building on this foundation, we investigate federated learning (FL) as a privacy-preserving approach to leverage collaboration across sites without centralizing data. Specifically, we evaluate the two top performing lightweight models from [5] (MobileNetV3Large and MobileNetV2), and additionally include MNv4-Conv-S (MobileNetV4-Conv-Small), which is designed for efficient on-device inference.

In healthcare, robust model development typically requires large and diverse datasets, yet clinical oral images are scarce, sensitive, and heavily regulated [25]. FL enables training across collaborators while keeping raw data local. This pilot study serves as a proof of concept and a practical blueprint for multicenter expansion: as additional institutions contribute locally held data, the proposed framework can scale to improve robustness and generalizability while preserving patient confidentiality.

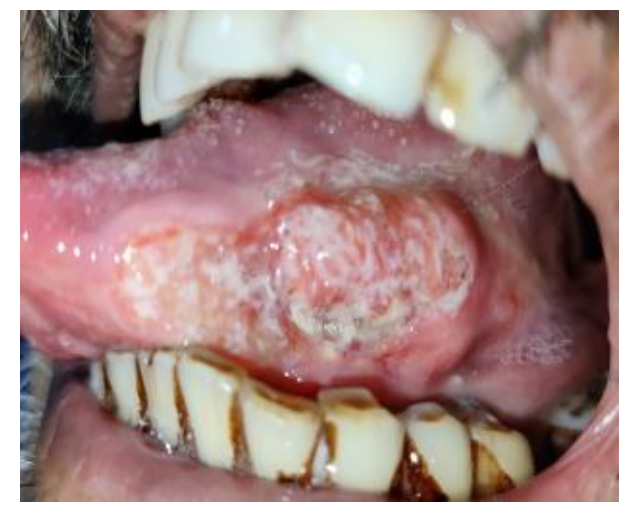
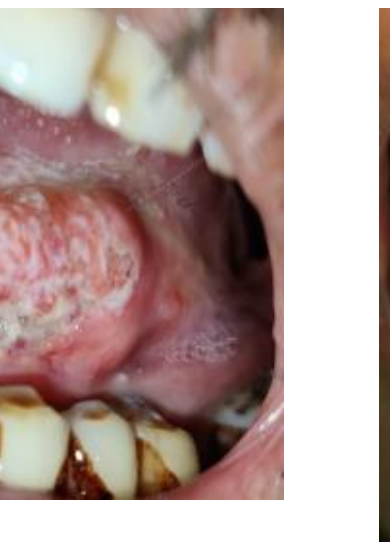

**(a) Cancer**

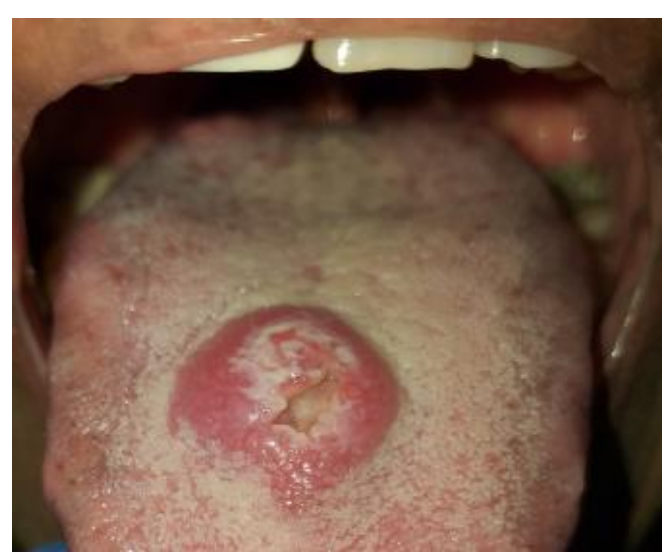

**(b) Cancer**

**(c) Normal**

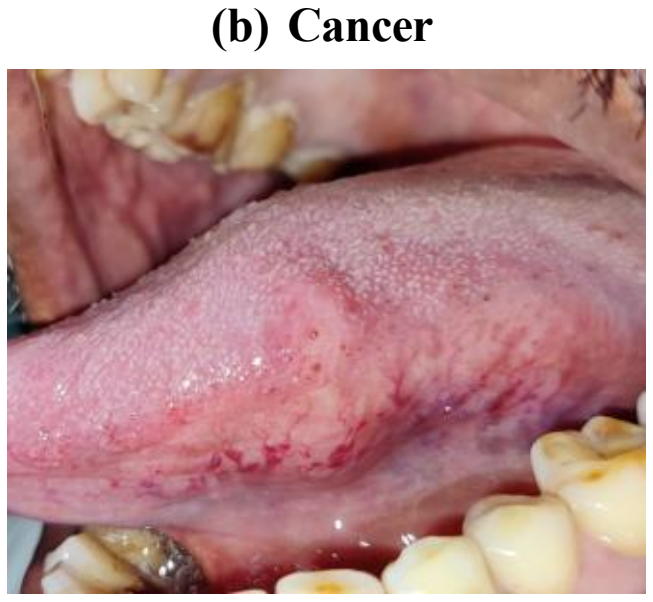

**(d) Normal**

Fig. 4. Sample images from the SmartOralpix dataset captured using smartphones.

### B. Datasets

**Client A (Dallas): SmartOralpix.** SmartOralpix is a privately collected dataset of clinical oral cavity photographs captured by dentists using smartphones (see Fig. 4). Each image was labeled by dentists and/or oral surgeons, providing clinically curated annotations for model training and evaluation.

**Client B (Philadelphia): Philly-Oral.** Client B's local dataset was constructed by pooling two sources to introduce feature variability across clients. The first source is a publicly available Mendeley dataset [28], consisting of 165 normal and 158 cancer images. For post-hoc calibration, 20 images per class were held out (40 total). The remaining 145 normal and 138 cancer images were used for Client B training. The second source is a private Temple University dataset extracted from the Kornberg School of Dentistry's axiUm® database, containing 18 normal and 2 cancer images. Combined, Client B's dataset contains 163 normal and 140 cancer images. We refer to this pooled dataset as *Philly-Oral* throughout the paper.

Although the Temple University subset is small, cancer-positive oral photographs are rare in routine acquisition; therefore, even a small number of confirmed cancer cases can be valuable for improving model coverage. The local dataset sizes are summarized in Table IV. Client A is perfectly balanced, and Client B exhibits only a small class imbalance (approximately 6%).

TABLE IV
CLIENT DATASETS USED FOR FEDERATED TRAINING

| Client | Location | Normal | Cancer | Total |
|---|---|---|---|---|
| A | Dallas | 148 | 148 | 296 |
| B | Philadelphia | 163 | 140 | 303 |

**Test set.** The test set was curated from a publicly available Kaggle dataset [29] and contains 10 normal and 21 cancer images. For local training, each client partitioned its dataset into 80% training and 20% validation. Despite the limited cohort size and the small number of clients, this configuration was sufficient to validate the end-to-end FL pipeline and demonstrate feasibility. The geographically separated client setup (Dallas and Philadelphia) further supports the practicality of cross-site collaboration for building models that are less dependent on acquisition conditions or local population characteristics.

### C. Data Preprocessing

We used MONAI (Medical Open Network for AI), a deep learning framework tailored for medical imaging, to implement preprocessing and data augmentation. Compared with general-purpose toolkits, MONAI provides healthcare-oriented transforms that simplify reproducible medical imaging pipelines.

To improve generalization under limited data, we applied on-the-fly data augmentation using MONAI transforms, including `LoadImageD`, `EnsureChannelFirstD`, `ScaleIntensityD`, `ResizeD`, `Lambdad`, `RandFlipD`, `ToTensorD`, and `Compose`. We excluded vertical flipping (spatial axis = 0) because it may distort clinically meaningful anatomical orientation.

### D. Federated Learning Framework

Federated learning can be executed either in a simulated environment (with virtual clients) or as cross-site training on independent client machines. Several open-source FL frameworks are available [24], including NVFlare, TensorFlow Federated (TFF), and FedML. We selected Flower [30] (Adap GmbH) for four reasons:

- **Cross-site support:** Flower supports both simulated and distributed cross-site training.
- **Research-oriented abstractions:** Flower provides clean interfaces for rapid experimentation and prototyping.
- **Model flexibility:** pretrained backbones can be integrated and customized with minimal overhead.
- **Framework interoperability:** Flower integrates naturally with PyTorch, TensorFlow, and Keras, and is well-suited to horizontal FL.

The proposed workflow (Fig. 5) was implemented using the Flower client interface for local training and the Flower server interface for coordination and aggregation.

### E. System Configuration

**Server.** Server-side coordination was executed on a 64-bit Windows laptop with an 8-core AMD Ryzen 7 4700U CPU and integrated Radeon graphics. The system ran Visual Studio, Python 3.9.5, and Flower v1.17.0.

**Client A (Dallas).** Client A experiments were conducted on a Mac Mini using Python 3.8.19, PyTorch v2.5.1, Torchvision v0.20.1, and Flower v1.18.0. Training and evaluation utilized Apple's Metal acceleration on an M-series chip (M1/M2) with 8 GB unified memory, and a 16-core Neural Engine.

**Client B (Philadelphia).** Client B experiments were conducted on a Windows desktop with an AMD Ryzen 9 7950X3D CPU, 64 GB RAM, and an NVIDIA GeForce RTX 4090 GPU. Software included PyTorch v2.5.1+cu124, Torchvision v0.20.1+cu124, Flower v1.18.0, and Python v3.12.9, using CUDA 12.4 for acceleration.

## IV. PROPOSED METHOD

### A. Secure Connectivity via Tailscale

We established cross-site communication for the proposed federated learning (FL) framework using Tailscale [31]. In our setup, Tailscale provides a secure mesh VPN (tailnet) that connects Client A (Dallas) and Client B (Philadelphia) to the central server, enabling real-time federated training across geographically separated sites. While SSH tunneling is a possible alternative, Tailscale offers a more stable, always-on connection after the initial device enrollment.

Each device was registered under its own Tailscale account. The server account then enabled device sharing by inviting each client device via its Tailscale-registered email address. After the client accepted the invitation, the device joined the server's shared tailnet. In this configuration, the server can view and connect to all shared client devices, whereas clients cannot see each other. This design keeps clients isolated and provides controlled, server-centric visibility.

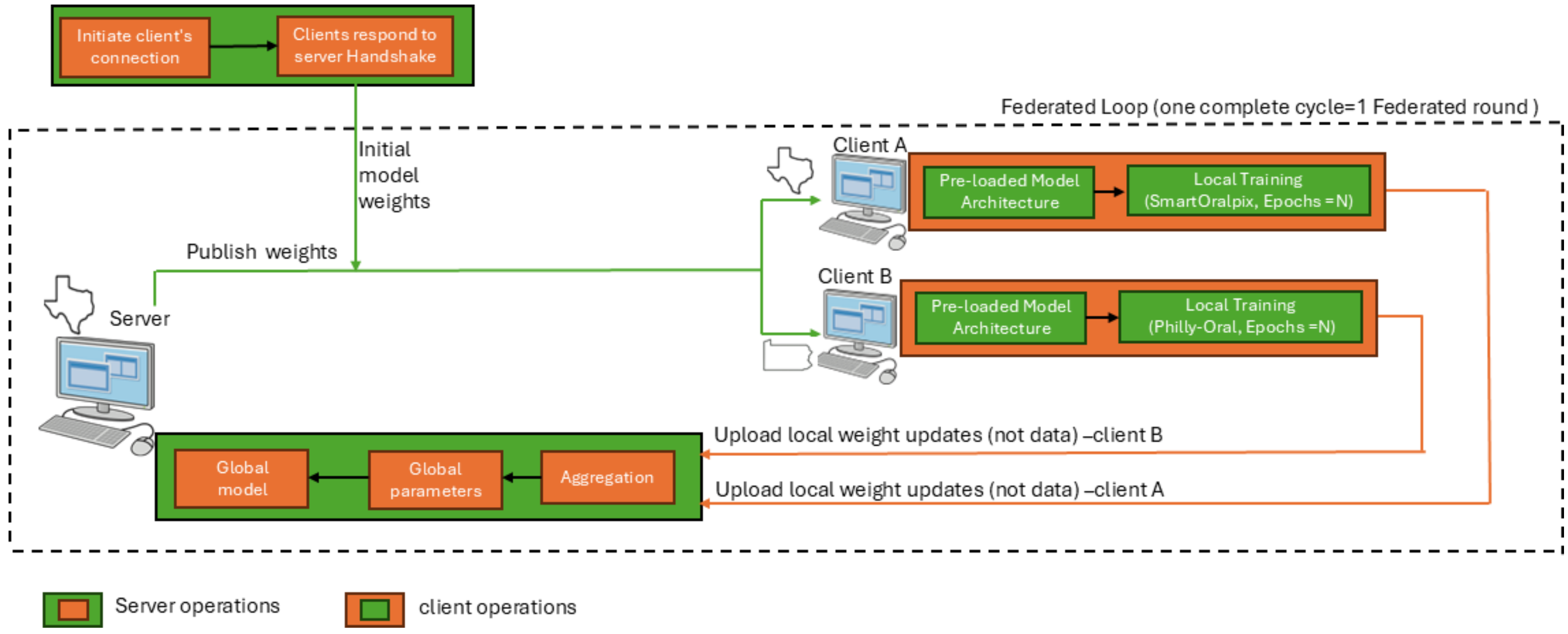

Fig. 5. Federated learning workflow used in this study. A secure Tailscale tunnel connects the central server and each client. The server initializes training by distributing model weights. Each client trains locally for a fixed number of epochs and returns updated weights. The server waits for all client updates, aggregates them into a global model (one federated round), and redistributes the updated weights. This process repeats until convergence.

For FL, each site runs a Flower client that connects to the Flower server using the server's Tailscale IP address. This ensures all communication occurs over the encrypted tailnet.

### B. Federated Learning Workflow

Fig. 5 illustrates the overall FL architecture. The system consists of a central server (Dallas) and two clients located at independent sites: Client A in Dallas and Client B in Philadelphia. The server orchestrates training, while each client performs local optimization on its private dataset.

Training proceeds as follows. The server performs a one-time initialization handshake and distributes the initial model parameters to all clients. Each client then trains locally for a fixed number of epochs and transmits updated model parameters to the server. The server waits for all clients to submit their updates, aggregates them into a new global model, and redistributes the updated parameters for the next cycle. This synchronous procedure repeats until convergence.

We use two terms throughout the paper:

- **Epoch:** one full pass over a client's local training dataset.
- **Round:** one full federated cycle consisting of server broadcast → local training → client upload → server aggregation.

### C. Client-Side Training

We adopt a homogeneous FL setting: all clients train the same model architecture, and the server aggregates updates for that same model. In Round 1, clients are initialized with ImageNet-1K pretrained weights distributed by the server. In each subsequent round, every client: (i) receives the current global weights, (ii) trains locally for the assigned number of epochs, and (iii) sends updated model parameters back to the server. This iterative process continues until the global model converges.

### D. Server-Side Aggregation and Global Selection

The central server coordinates all rounds and performs model aggregation. Among the many aggregation strategies available in FL, we use Federated Averaging (FedAvg) [32] for all lightweight backbones evaluated in this study. In each round, the server receives model updates from Client A and Client B, aggregates them, and produces the updated global model.

Because the server does not hold a dedicated global validation dataset, we estimate global validation performance using the clients' local validation metrics. Specifically, we compute weighted averages of validation loss and validation accuracy, where weights correspond to the number of validation samples contributed by each client.

$$Loss_{\mathrm{global}} = \frac{n_A\, Loss_A + n_B\, Loss_B}{n_A + n_B} \tag{1}$$

$$Accuracy_{\mathrm{global}} = \frac{n_A\, Accuracy_A + n_B\, Accuracy_B}{n_A + n_B} \tag{2}$$

Here, $n_A$ and $n_B$ denote the number of validation samples at Client A and Client B, respectively. In Table V, the "global metrics" column is computed using these expressions. We select the best global checkpoint based on the highest $Accuracy_{\mathrm{global}}$. The corresponding global weights are saved as a `.pth` file and used for test-set evaluation.

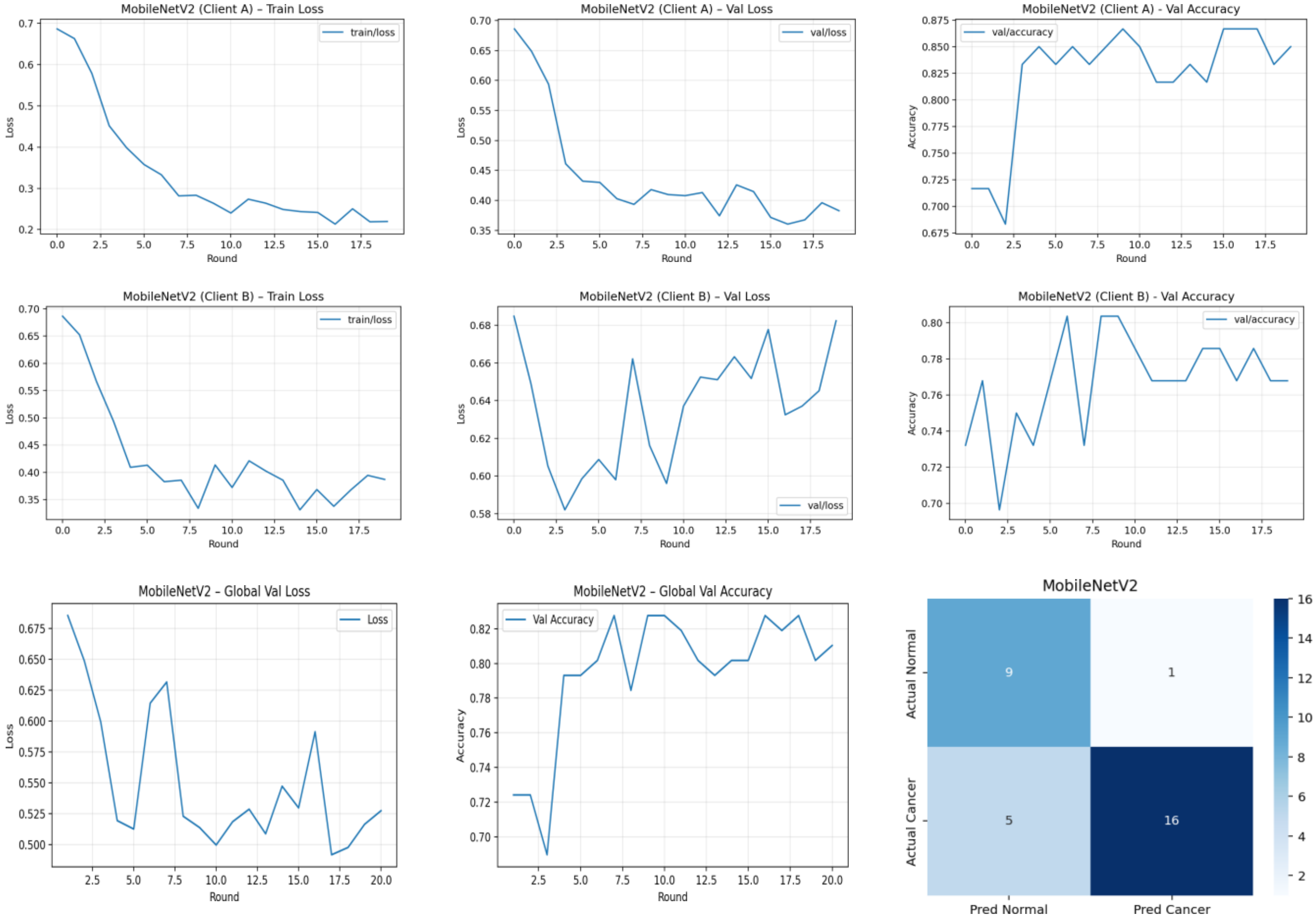

Fig. 6. MobileNetV2 performance under FL. Top row: Client A training loss, validation loss, and validation accuracy on SmartOralpix. Second row: Client B training loss, validation loss, and validation accuracy on Philly-Oral. Bottom row: GM-V2 (global MobileNetV2) global validation loss, global validation accuracy, and confusion matrix on the test set.

### *E. Final Global Model and Calibration*

We denote the final aggregated global models as GM-V2 (MobileNetV2), GM-V3L (MobileNetV3Large), and GM-V4S (MNv4-Conv-S). Modern neural networks are often poorly calibrated [33], which is particularly problematic in medical AI where confidence scores can affect downstream decisions. We therefore apply post-hoc calibration consisting of BatchNorm (BN) warm-up followed by temperature scaling.

In the federated setting, calibration is performed centrally on the selected global model using a held-out calibration set comprising 20 normal and 20 cancer images from the Mendeley dataset.

*1) BatchNorm Warm-up:* FedAvg aggregates model parameters but typically does not aggregate BatchNorm running statistics (buffers). Since each client observes a different data distribution, BN statistics can drift differently across clients. We therefore perform BN warm-up on the global model prior to evaluation by forwarding the calibration dataset through the model in training mode (without weight updates). This updates BN running mean and variance to better reflect the calibration distribution.

*2) Temperature Scaling:* After BN warm-up, the logit may still yield overconfident or underconfident probabilities. Temperature scaling rescales logit by a single scalar $T$ to improve probabilistic calibration without changing the ranking of predictions. As a result, discrimination metrics such as AUC are unchanged, while probability estimates become better aligned with empirical outcomes.

$$\hat{p} = \sigma\left(\frac{z}{T}\right) \tag{3}$$

Here, $\hat{p}$ denotes the calibrated predicted probability, $\sigma(\cdot)$ is the sigmoid activation function, $z$ represents the model logit (raw output before activation), and $T > 0$ is the temperature parameter used to adjust the confidence of the predictions.

### *F. Core ML Packaging for iOS Deployment*

The best-performing model (GM-V4S) is initialized and loaded with the saved PyTorch checkpoint (`.pth`). The model is then converted to TorchScript and packaged as a Core ML model (`.mlpackage`) for direct integration into the iOS oral cancer screening application.

## V. Results and Discussion

This section analyzes (i) client-side training dynamics and (ii) the performance of the aggregated global models. Client training loss reflects the stability of optimization on each local dataset. Client validation loss indicates local generalization, while validation accuracy measures local discriminative performance. The global validation loss and accuracy curves in Figs. 6, 7, and 8 summarize aggregated performance across clients and are computed from the client's local validation metrics using (1)–(2).

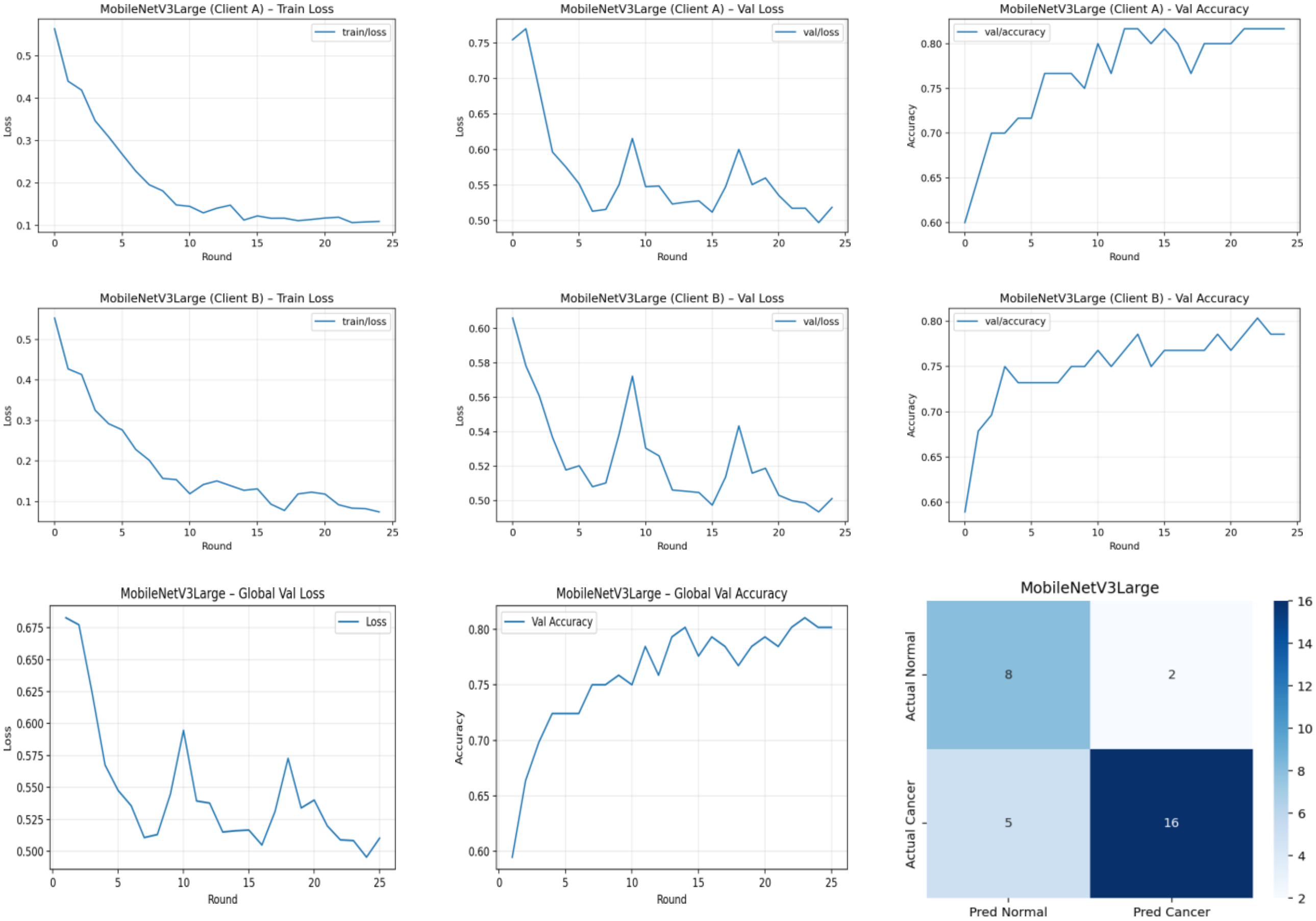

Fig. 7. MobileNetV3Large performance under FL. Top row: Client A training loss, validation loss, and validation accuracy on SmartOralpix. Second row: Client B training loss, validation loss, and validation accuracy on Philly-Oral. Bottom row: GM-V3L (global MobileNetV3Large) global validation loss, global validation accuracy, and confusion matrix on the test set.

### A. Comparison of Local Client Performance

We adopted a consistent transfer-learning strategy across all models. During the initial rounds, the backbone was frozen (backbone learning rate $= 0$) and only the classifier head was trained. Each federated round used three local epochs in this phase. After unfreezing, we fine-tuned the full network using separate learning rates for the backbone and the classifier head, with the head learning rate set substantially higher (typically at least $0.3\times$ higher) to avoid destabilizing pretrained features. To account for cross-site heterogeneity, learning rate and weight decay were allowed to differ across clients. In the final rounds, we reduced local epochs to two and used small learning rates ($2 \times 10^{-5}$ and $4 \times 10^{-5}$) to stabilize global aggregation.

This procedure served as the baseline across experiments. While the number of federated rounds and local epochs was kept consistent across clients, hyperparameters were tuned per model. The final configuration for each model was selected based on the highest global validation accuracy observed during training.

*1) MobileNetV2:* Fig. 6 shows MobileNetV2 behavior under FL. Client A exhibits a smooth, near-monotonic decrease in training loss, indicating stable local optimization and correspondingly steady improvement in validation loss and accuracy. Client B also shows decreasing training loss, but with noticeable oscillations, consistent with less stable optimization under heterogeneous data. This behavior is reflected in a more variable validation loss and lower validation accuracy at Client B. As a result, the global validation loss curve contains intermittent spikes, and the global validation accuracy fluctuates moderately. The model was trained for 20 rounds to reach convergence.

*2) MobileNetV3Large:* Fig. 7 shows MobileNetV3Large trained for 25 rounds. Both clients demonstrate stable convergence with decreasing training loss. However, Client A exhibits more stable validation behavior and higher validation accuracy than Client B, suggesting better generalization under cross-client variability. Compared with MobileNetV2, the global validation curves are smoother but do not translate into superior test performance.

*3) MNv4-Conv-S:* Fig. 8 shows MNv4-Conv-S trained for 20 rounds. Both clients exhibit a steep early reduction in training loss followed by a gradual decrease, indicating rapid initial learning within the first few rounds. Client A maintains relatively stable validation loss, whereas Client B shows larger fluctuations, suggesting weaker local generalization. Client A also attains slightly higher validation accuracy, consistent with its steadier validation loss.

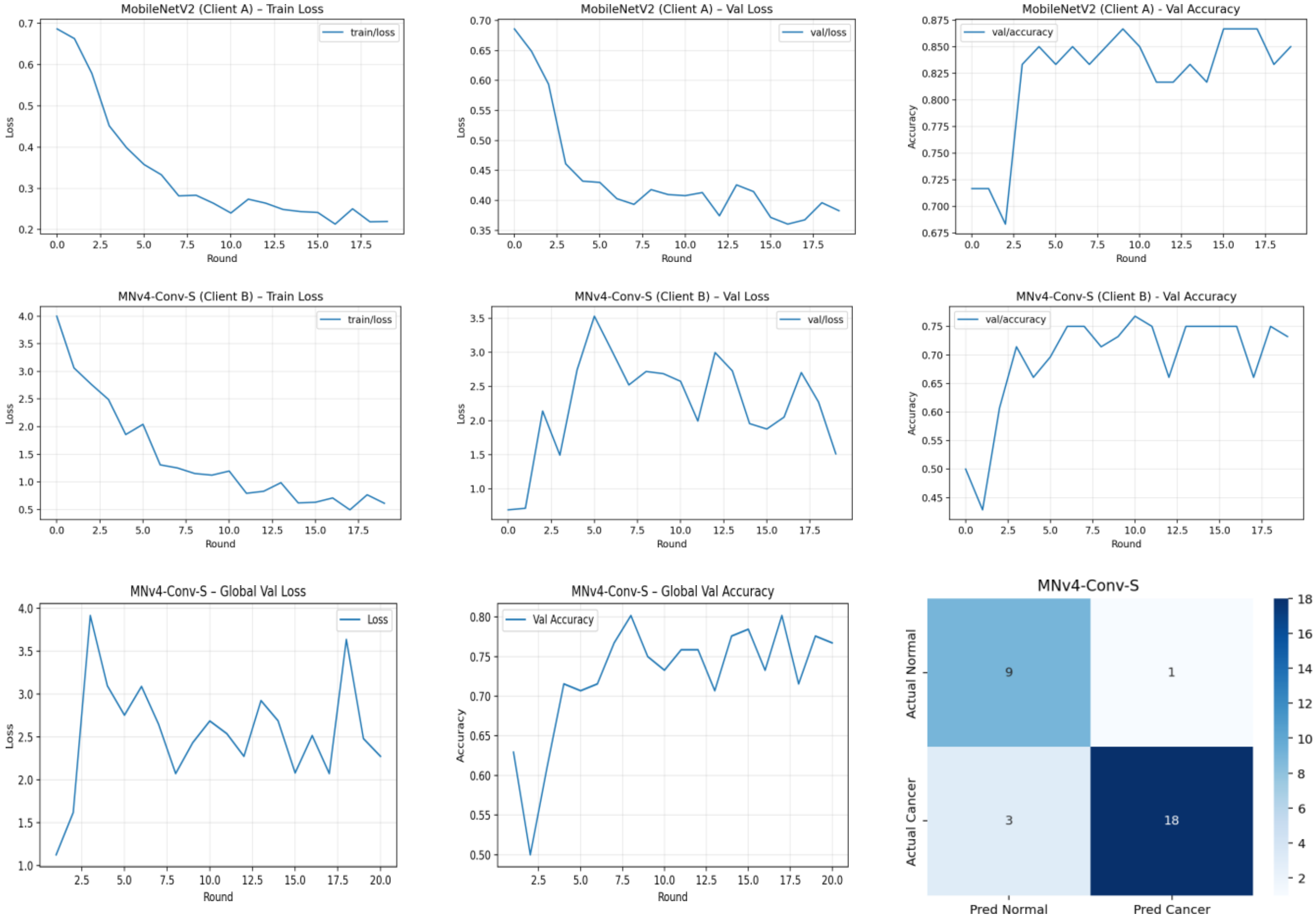

Fig. 8. MNv4-Conv-S performance under FL. Top row: Client A training loss, validation loss, and validation accuracy on SmartOralpix. Second row: Client B training loss, validation loss, and validation accuracy on Philly-Oral. Bottom row: GM-V4S (global MNv4-Conv-S) global validation loss, global validation accuracy, and confusion matrix on the test set.

Table V reports the best checkpoint for each model (selected by highest global validation accuracy). Across all backbones, Client B shows higher validation loss and lower validation accuracy than Client A, indicating greater difficulty generalizing—likely due to cross-site heterogeneity and the pooled composition of the Philly-Oral dataset.

TABLE V
BEST-CHECKPOINT VALIDATION PERFORMANCE DURING FEDERATED TRAINING

| Model | Metric | Client A | Client B | Global |
|---|---|---|---|---|
| MobileNetV2 | Loss | 0.4098 | 0.5961 | 0.4997 |
| | Accuracy | 0.8667 | 0.8036 | 0.8362 |
| MobileNetV3Large | Loss | 0.5176 | 0.4985 | 0.5083 |
| | Accuracy | 0.8167 | 0.8036 | 0.8103 |
| MNv4-Conv-S | Loss | 1.6510 | 2.5232 | 2.0721 |
| | Accuracy | 0.8499 | 0.7500 | 0.8017 |

## *B. Test-Set Performance of Global Models*

We evaluated the three aggregated global models on the held-out test set. Table VI reports sensitivity, specificity, F1 score, accuracy, and AUC. GM-V4S achieves the best overall performance and notably improves sensitivity (85.71%), which is critical in screening workflows where false negatives carry high clinical risk. GM-V4S also achieves the highest accuracy (87.10%) and AUC (0.929), indicating strong discriminative performance.

## *C. Discussion*

The bottom rows of Figs. 6, 7, and 8 summarize global validation behavior and show the confusion matrices for GM-V2, GM-V3L, and GM-V4S on the test set. Among the three, GM-V4S yields the strongest confusion matrix profile, with improved cancer detection and fewer missed cancer cases relative to the other models.

As shown in Table VI and Fig. 9, GM-V4S provides the best trade-off across metrics, and its AUC of 0.929 indicates superior separation between normal and cancer classes. This makes GM-V4S the most suitable candidate for a screening-oriented deployment in our federated setting.

We initially expected MobileNetV2 and MobileNetV3Large to perform similarly under FL, consistent with their centralized results in [5]. However, the federated results indicate that strong centralized performance does not necessarily transfer to equivalent FL performance. One plausible contributor is the behavior of BatchNorm (BN). MobileNetV3Large relies heavily on BN layers, while standard FedAvg aggregates model parameters but does not explicitly reconcile BN running statistics across heterogeneous clients. This mismatch can degrade generalization when client data distributions differ, potentially explaining the reduced test performance of GM-V3L relative to GM-V2 and, in particular, GM-V4S.

TABLE VI
TEST-SET PERFORMANCE OF GLOBAL MODELS

| Global Model | Sensitivity(%) | Specificity(%) | F1 score(%) | Accuracy(%) | AUC |
|---|---|---|---|---|---|
| GM-V2 | 76.19 | 90 | 84.21 | 80.65 | 0.833 |
| GM-V3L | 76.19 | 80 | 82.05 | 77.42 | 0.819 |
| **GM-V4S** | **85.71** | **90** | **90** | **87.10** | **0.929** |

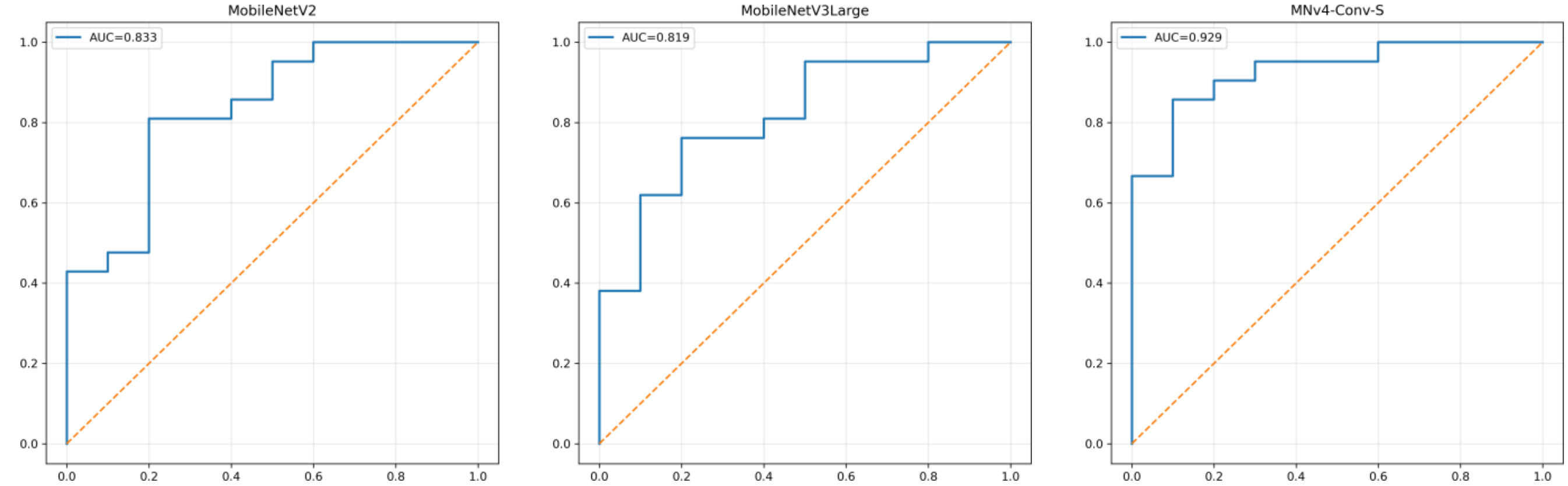

Fig. 9. ROC curves for GM-V2, GM-V3L, and GM-V4S on the test set.

## VI. MOBILE APPLICATION

We developed an iOS application that performs on-device oral cancer screening and integrated the best-performing global model (GM-V4S) as a Core ML package. The application includes a lightweight user interface for image capture and selection, followed by local inference executed entirely on the smartphone. Performing inference on-device reduces latency, avoids network dependency, and eliminates the need to transmit patient images off the device.

The model output is presented using user-friendly labels to support screening-oriented use: predictions classified as cancer are displayed as *Suspicious*, while non-cancer predictions are displayed as *Non-Suspicious*. In this study, the app is used as a proof of concept for mobile deployment and real-time inference. Model quantization and additional runtime optimizations (e.g., pruning and hardware-aware compression) were not applied in the current phase and are reserved for future work.

## VII. CONCLUSIONS

This paper presented a practical federated learning (FL) framework for training privacy-preserving oral cancer screening models using decentralized clinical oral photographs. By enabling cross-site collaboration without centralizing raw patient data, the proposed workflow supports scalable model development while maintaining data locality and confidentiality. The framework is lightweight and reproducible, and can be readily adapted by researchers to other healthcare applications requiring privacy-aware collaboration.

Our results highlight two key observations. First, careful backbone selection is critical: choosing a model architecture that matches the data modality and deployment constraints provides a strong foundation for representation learning before exploring more sophisticated aggregation strategies. Second, model behavior under centralized training does not necessarily predict performance under FL, reinforcing the importance of evaluating candidate backbones directly in federated settings. From an application perspective, the integrated iOS prototype demonstrates the feasibility of deploying the best global model for fully on-device inference, enabling a teledentistry-oriented triage tool that can support early screening in remote or resource-limited contexts. Overall, this pilot study establishes a concrete blueprint for multi-institution collaboration in oral cancer AI and motivates broader, privacy-preserving dataset expansion.

Future work will evaluate additional aggregation and personalization methods beyond FedAvg, and will extend collaboration to more institutions to increase the number of participating clients and the diversity of data. We also plan to investigate model compression and quantization-aware training within the FL workflow, to further optimize runtime, memory footprint, and energy consumption for mobile deployment.

## ACKNOWLEDGMENT

Lakshman Tamil gratefully acknowledges the inspiration provided by the late Mr. Gregory Stubblefield and dedicates this paper to his memory. ChatGPT-4.0 was used to assist with English-language editing in portions of the manuscript.

## References

[1] American Cancer Society, "Key statistics for oral cavity and oropharyngeal cancers," https://www.cancer.org/cancer/types/oral-cavity-and-oropharyngeal-cancer/about/key-statistics.html, 2026, accessed: 2026-01-21.

[2] R. Sankaranarayanan, K. Ramadas, G. Thomas, R. Muwonge, S. Thara, B. Mathew, and B. Rajan, "Effect of screening on oral cancer mortality in kerala, india: a cluster-randomised controlled trial," *The Lancet*, vol. 365, no. 9475, pp. 1927–1933, 2005.

[3] Surveillance Research Program, National Cancer Institute, "SEER*Explorer: An interactive website for SEER cancer statistics," Jan. 2026, internet. Published 2025-04-16; updated 2026-01-08; cited 2026-01-27. Data source(s): SEER Incidence Data, November 2024 Submission (1975–2022), SEER 21 registries. [Online]. Available: https://seer.cancer.gov/statistics-network/explorer/

[4] U.S. Department of Health and Human Services, "Hippa," https://www.hhs.gov/hipaa/index.html, 1996, accessed: 2025-01-05.

[5] L. D. Swamikannan, A. B. Sonawane, J. S. Patel, C. Mani, L. Narayana, and L. Tamil, "Oral cancer detection using mobile vision technology," in *2024 IEEE EMBS International Conference on Biomedical and Health Informatics (BHI)*, 2024, pp. 1–8.

[6] S. Camalan, H. Mahmood, H. Binol, A. L. D. Araujo, A. R. Santos-Silva, P. A. Vargas, M. A. Lopes, S. A. Khurram, and M. N. Gurcan, "Convolutional neural network-based clinical predictors of oral dysplasia: class activation map analysis of deep learning results," *Cancers*, vol. 13, no. 6, p. 1291, 2021.

[7] K. Warin, W. Limprasert, S. Suebnukarn, S. Jinaporntham, and P. Jantana, "Automatic classification and detection of oral cancer in photographic images using deep learning algorithms," *Journal of Oral Pathology & Medicine*, vol. 50, no. 9, pp. 911–918, 2021.

[8] K. Bansal, R. Bathla, and Y. Kumar, "Deep transfer learning techniques with hybrid optimization in early prediction and diagnosis of different types of oral cancer," *Soft Computing*, vol. 26, no. 21, pp. 11 153–11 184, 2022.

[9] R. A. Welikala, P. Remagnino, J. H. Lim, C. S. Chan, S. Rajendran, T. G. Kallarakkal, R. B. Zain, R. D. Jayasinghe, J. Rimal, A. R. Kerr *et al.*, "Automated detection and classification of oral lesions using deep learning for early detection of oral cancer," *IEEE Access*, vol. 8, pp. 132 677–132 693, 2020.

[10] F. Jubair, O. Al-karadsheh, D. Malamos, S. Al Mahdi, Y. Saad, and Y. Hassona, "A novel lightweight deep convolutional neural network for early detection of oral cancer," *Oral Diseases*, vol. 28, no. 4, pp. 1123–1130, 2022.

[11] D. Sharma, V. Kudva, V. Patil, A. Kudva, and R. S. Bhat, "A convolutional neural network based deep learning algorithm for identification of oral precancerous and cancerous lesion and differentiation from normal mucosa: a retrospective study," *Engineered Science*, vol. 18, no. 15, pp. 278–287, 2022.

[12] T. Flu¨gge, R. Gaudin, A. Sabatakakis, D. Tro¨ltzsch, M. Heiland, N. van Nistelrooij, and S. Vinayahalingam, "Detection of oral squamous cell carcinoma in clinical photographs using a vision transformer," *Scientific Reports*, vol. 13, no. 1, p. 2296, 2023.

[13] N. Firdaus and Z. Raza, "Enhancing privacy in oral cancer detection through federated learning: A cross-institutional study," *Procedia Computer Science*, vol. 260, pp. 1113–1120, 2025.

[14] M. Sandler, A. Howard, M. Zhu, A. Zhmoginov, and L.-C. Chen, "Mobilenetv2: Inverted residuals and linear bottlenecks," in *Proceedings of the IEEE Conference on Computer Vision and Pattern Recognition (CVPR)*, June 2018.

[15] A. Howard, R. Pang, H. Adam, Q. Le, M. Sandler, B. Chen, W. Wang, L.-C. Chen, M. Tan, G. Chu, V. Vasudevan, and Y. Zhu, "Searching for mobilenetv3," 10 2019, pp. 1314–1324.

[16] D. Qin, C. Leichner, M. Delakis, M. Fornoni, S. Luo, F. Yang, W. Wang, C. Banbury, C. Ye, B. Akin *et al.*, "Mobilenetv4: Universal models for the mobile ecosystem," in *European Conference on Computer Vision*. Springer, 2024, pp. 78–96.

[17] A.-u. Rahman, A. Alqahtani, N. Aldhafferi, M. U. Nasir, M. F. Khan, M. A. Khan, and A. Mosavi, "Histopathologic oral cancer prediction using oral squamous cell carcinoma biopsy empowered with transfer learning," *Sensors*, vol. 22, no. 10, p. 3833, 2022.

[18] N. Haj-Hosseini, J. Lindblad, B. Hasse´us, V. V. Kumar, N. Subramaniam, and J.-M. Hirsch, "Early detection of oral potentially malignant disorders: a review on prospective screening methods with regard to global challenges," *Journal of Maxillofacial and Oral Surgery*, vol. 23, no. 1, pp. 23–32, 2024.

[19] K. Bansal, R. Bathla, and Y. Kumar, "Deep transfer learning techniques with hybrid optimization in early prediction and diagnosis of different types of oral cancer," *Soft Computing*, vol. 26, no. 21, pp. 11 153–11 184, 2022.

[20] J. Folmsbee, X. Liu, M. Brandwein-Weber, and S. Doyle, "Active deep learning: Improved training efficiency of convolutional neural networks for tissue classification in oral cavity cancer," in *2018 IEEE 15th international symposium on biomedical imaging (ISBI 2018)*. IEEE, 2018, pp. 770–773.

[21] B. Singha Deo, M. Pal, P. K. Panigrahi, and A. Pradhan, "Supremacy of attention based convolution neural network in classification of oral cancer using histopathological images," *medRxiv*, pp. 2022–11, 2022.

[22] L. Peng, G. Luo, A. Walker, Z. Zaiman, E. K. Jones, H. Gupta, K. Kersten, J. L. Burns, C. A. Harle, T. Magoc *et al.*, "Evaluation of federated learning variations for covid-19 diagnosis using chest radiographs from 42 us and european hospitals," *Journal of the American Medical Informatics Association*, vol. 30, no. 1, pp. 54–63, 2022.

[23] S. Rajit and M. A. Al Sayed, "Federated learning based histopathological image classification for oral squamous cell carcinoma," in *2024 IEEE-EMBS Conference on Biomedical Engineering and Sciences (IECBES)*. IEEE, 2024, pp. 339–344.

[24] B. Yurdem, M. Kuzlu, M. K. Gullu, F. O. Catak, and M. Tabassum, "Federated learning: Overview, strategies, applications, tools and future directions," *Heliyon*, 2024.

[25] H. Li, C. Li, J. Wang, A. Yang, Z. Ma, Z. Zhang, and D. Hua, "Review on security of federated learning and its application in healthcare," *Future Generation Computer Systems*, vol. 144, pp. 271–290, 2023.

[26] Y. Tan, G. Long, L. Liu, T. Zhou, Q. Lu, J. Jiang, and C. Zhang, "Fedproto: Federated prototype learning across heterogeneous clients," in *Proceedings of the AAAI conference on artificial intelligence*, vol. 36, no. 8, 2022, pp. 8432–8440.

[27] N. Saeed, M. Ashour, and M. Mashaly, "Comprehensive review of federated learning challenges: a data preparation viewpoint," *Journal of Big Data*, vol. 12, no. 1, p. 153, 2025.

[28] H. S. Chandrashekar, A. Geetha Kiran, S. Murali, M. S. Dinesh, and B. R. Nanditha, "Oral images dataset," in *Mendeley Data*, 2021. [Online]. Available: https://doi.org/10.17632/mhjyrn35p4.2

[29] S. Barot and P. Suthar, "Oral cancer (lips and tongue) images," Kaggle, 2020, [Online]. [Online]. Available: https://www.kaggle.com/datasets/sbarot/oral-cancer-lips-and-tongue-images

[30] D. J. Beutel, T. Topal, A. Mathur, X. Qiu, J. Fernandez-Marques, Y. Gao, L. Sani, K. H. Li, T. Parcollet, P. P. B. de Gusma˜o *et al.*, "Flower: A friendly federated learning research framework," *arXiv preprint arXiv:2007.14390*, 2020.

[31] Tailscale Inc., "Tailscale," https://tailscale.com, 2025, version 1.84.2.

[32] B. McMahan, E. Moore, D. Ramage, S. Hampson, and B. A. y Arcas, "Communication-efficient learning of deep networks from decentralized data," in *Artificial intelligence and statistics*. PMLR, 2017, pp. 1273–1282.

[33] C. Guo, G. Pleiss, Y. Sun, and K. Q. Weinberger, "On calibration of modern neural networks," in *International conference on machine learning*. PMLR, 2017, pp. 1321–1330.